\documentclass[11pt,letterpaper]{article}
\usepackage[numbers]{natbib}
\usepackage[margin=.75in]{geometry}
\usepackage{mathpazo,color}
\usepackage{siunitx}
\usepackage{graphicx}%
\usepackage{epstopdf}
\usepackage{hyperref}
\usepackage[fleqn]{amsmath}
\usepackage[referable]{threeparttablex}
\usepackage{multirow}
\usepackage{enumitem,booktabs,cfr-lm}
\usepackage[referable]{threeparttablex}
\usepackage{color}
\definecolor{dark-gray}{gray}{0.4}

\usepackage{float}

\newcommand{\sumjd}{\sum_{j=1}^d}
\newcommand{\sumin}{\sum_{i=1}^n}
\newcommand{\bbeta}{\boldsymbol{\beta}}

\newcommand{\blambda}{\boldsymbol{\lambda}}
\newcommand{\blambdahat}{\widehat{\boldsymbol{\lambda}}}

\newcommand{\btheta}{\boldsymbol{\theta}}
\newcommand{\bthetahat}{\widehat{\boldsymbol{\theta}}}
\newcommand{\bthetak}{\boldsymbol{\theta}^{(k)}}
\newcommand{\bO}{\boldsymbol{O}}
\newcommand{\bz}{\boldsymbol{z}}
\newcommand{\bZ}{\boldsymbol{Z}}

\newcommand{\wijk}{w_{ij}^{(k)}}
\newcommand{\wplusjk}{w_{+j}^{(k)}}
\newcommand{\wiplusk}{w_{i+}^{(k)}}

\newcommand{\betaem}{\widehat{\beta}_{em}}

\newcommand{\betaw}{\widehat{\beta}_{w}}
\newcommand{\betas}{\widehat{\beta}_{s}}

\newcommand{\betaemone}{\widehat{\beta}_{em,1}}
\newcommand{\betaemtwo}{\widehat{\beta}_{em,2}}
\newcommand{\betawone}{\widehat{\beta}_{w,1}}
\newcommand{\betawtwo}{\widehat{\beta}_{w,2}}
\newcommand{\betasone}{\widehat{\beta}_{s,1}}
\newcommand{\betastwo}{\widehat{\beta}_{s,2}}

\newcommand{\betaemj}{\widehat{\beta}_{em,j}}
\newcommand{\betawj}{\widehat{\beta}_{w,j}}
\newcommand{\betasj}{\widehat{\beta}_{s,j}}

\newcommand{\bbetaem}{\widehat{\boldsymbol{\beta}}_{em}}
\newcommand{\bbetaw}{\widehat{\boldsymbol{\beta}}_w}
\newcommand{\bbetas}{\widehat{\boldsymbol{\beta}}_s}
\newcommand{\bbetahat}{\widehat{\boldsymbol{\beta}}}

\newcommand{\Lambdaemhat}{\widehat{\Lambda}_{em}}
\newcommand{\Lambdahatn}{\widehat{\Lambda}_{n}}
\newcommand{\bbetahatn}{\widehat{\boldsymbol{\beta}}_{n}}
\newcommand{\bthetahatn}{\widehat{\boldsymbol{\theta}}_n}

\newcommand{\lntheta}{l_n(\btheta)}
\newcommand{\lnthetahat}{l_n(\bthetahat_n)}
\newcommand{\lnthetazero}{l_n(\btheta_0)}

\renewlist{tablenotes}{enumerate}{1}
\makeatletter
\setlist[tablenotes]{label=\tnote{\alph*},ref=\alph*,itemsep=\z@,topsep=\z@skip,partopsep=\z@skip,parsep=\z@,itemindent=\z@,labelindent=\tabcolsep,labelsep=.2em,leftmargin=*,align=left,before={\footnotesize}}
\makeatother

\title{\textbf{Cox Regression Model under Dependent Truncation}}
\author{\textbf{Lior Rennert and Sharon X. Xie} \\ Department of Biostatistics Epidemiology and Informatics, University of Pennsylvania, 110 Blockley Hall, \\423 Guardian Drive, Philadelphia, Pennsylvania 19104, U.S.A.}
\date{}
\begin{document}
\maketitle

\begin{abstract}
Truncation is a statistical phenomenon that occurs in many time to event studies. For example, autopsy-confirmed studies of neurodegenerative diseases are subject to an inherent left and right truncation, also known as double truncation. When the goal is to study the effect of risk factors on survival, the standard Cox regression model cannot be used when the data is subject to truncation. Existing methods which adjust for both left and right truncation in the Cox regression model require independence between the survival times and truncation times, which may not be a reasonable assumption in practice. We propose an expectation-maximization algorithm to relax the independence assumption in the Cox regression model under left, right, or double truncation, to an assumption of conditional independence. The resulting regression coefficient estimators are consistent and asymptotically normal. We demonstrate through extensive simulations that the proposed estimators have little bias and, in most practical situations, have a lower mean-squared error compared to existing estimators. We implement our approach to assess the effect of occupation on survival in subjects with autopsy-confirmed Alzheimer's disease.
\end{abstract}

\section{Introduction}
Truncation is a statistical phenomenon that has been shown to occur in a wide range of applications, including survival analysis, epidemiology, economics, and astronomy. Individuals who are subject to truncation provide no information to the investigator. \textit{Left truncation} occurs when data is only recorded for individuals whose event time exceeds a random time (i.e. left truncation time). Under left truncation, individuals with smaller event times are less likely to be observed, resulting in a study sample that is biased towards larger event times and risk factors associated with larger event times. \textit{Right truncation} occurs when data is only recorded for individuals whose event time proceeds a random time (i.e. right truncation time). Under right truncation, individuals with larger event times are less likely to be observed, resulting in a study sample that is biased towards smaller event times and risk factors associated with smaller event times. When both left and right truncation are present, this is known as \textit{double truncation}.

Double truncation is inherent in autopsy-confirmed studies of neurodegenerative diseases \cite{rennert_cox_2017}. Left truncation occurs because individuals enter the study after the onset of the disease, and therefore those who succumb to the disease before they enter the study are unobserved. The right truncation occurs because individuals who live past the end of the study date do not receive a pathological diagnosis of the disease. Since these subjects cannot be definitively diagnosed with a particular disease, they are excluded from the autopsy-confirmed study sample and therefore provide no information to the investigator. This is contrary to censored individuals, who provide partial information about their survival time. We note, however, that right censoring is not possible in autopsy-confirmed studies, since any individual who has an autopsy performed will also have a known survival time. This truncation scheme is illustrated in Figure 1, where only individuals whose time of death falls between the study entry time and end of study time are observed. 

The aim of our data analysis is to get accurate estimates of the effect of risk factors on survival from disease symptom onset in subjects with autopsy-confirmed Alzheimer's disease (AD), the most common neurodegenerative disease. Because individuals with shorter survival times are less likely to enter the study, left truncation leads to a study sample that is biased towards larger survival times and risk factors associated with larger survival times. Similarly, individuals with longer survival times are more likely to live past the end of the study, and therefore right truncation leads to a study sample that is biased towards smaller survival times and risk factors associated with smaller survival times. If double truncation is not accounted for, then the regression coefficient estimators from the Cox regression model \cite{cox_regression_1972} will be biased.

Methods to handle double truncation have recently started gaining traction in the literature. In 2017, three methods were published to adjust the Cox model under double truncation \citep{rennert_cox_2017,shen_pseudo_2017,mandel_inverse_2017}. The estimation procedure for all three methods rely on estimating the joint distribution of the left and right truncation times, which is used to compute the probability that a subject is observed (i.e. not truncated). These probabilities are then used as weights or offsets in the Cox model. However, the estimation of the truncation distribution relies on the assumption of independence between the observed survival and truncation times, which may not a reasonable assumption in practice. For example, according to the Alzheimer's association and discussions with our clinical investigators, factors such as lower age of symptom onset, depression, and stress are associated with delayed study entry. Since these factors are associated with survival, this induces a dependence between the left truncation times and survival times. As shown in the simulation studies in Section 3, the regression coefficient estimators from \citep{rennert_cox_2017,shen_pseudo_2017,mandel_inverse_2017} are sensitive to violations of this independence assumption. Therefore, the existing literature is unable to address the unique challenges present in our study.  

In this paper, we propose a novel method to relax the assumption of independence between the observed survival and truncation times in the Cox proportional hazards model under left, right, and double truncation. Specifically, by conditioning on the observed truncation times, our method relaxes the independence assumption to an assumption of \textit{conditional independence}. Treating the truncated survival times as missing, we introduce an expectation-maximization (EM) algorithm to estimate the regression coefficients and baseline hazard rates. This approach, which completely avoids the estimation of the truncation distribution, yields consistent and asymptotically normal regression coefficient estimators. We show through extensive simulation studies that our proposed estimators have little bias in small samples, while the estimators based on the methods introduced in \citep{rennert_cox_2017,shen_pseudo_2017,mandel_inverse_2017}, and the standard model which ignores double truncation, can be heavily biased under violations of the independence assumption. We show that even if the independence assumption is satisfied, our proposed method performs as well as the existing approaches. We illustrate our method by analyzing the effect of cognitive reserve on survival in an autopsy-confirmed AD cohort.

Cognitive reserve (CR) is a widely used hypothetical construct intended to account for individual differences in cognitive decline and clinical manifestations of dementia among individuals with AD \citep{Stern_Cognitive_2012,Meng_Education_2012}. CR hypothesizes that individuals develop cognitive strategies and neural connections throughout their life times through experience such as occupation, education, and other forms of mental engagement \citep{Valenzuela_Assessment_2007}. This may modulate the effects of AD because of compensatory strategies obtained from a higher level of professional performance or a good education \citep{Sanchez_Study_2011}. For example, CR may have a protective role in the brain and therefore lengthen survival from disease symptom onset \citep{Ientile_Survival_2013}. 

Occupation, often used as a proxy for CR, has been shown to modulate survival in healthy aging and AD \citep{Massimo_Occupational_2015}. Several studies in healthy aging have examined the possible protective influence of higher occupational attainment on survival \citep{enroth_is_2014,Andel_role_2014,correa_ribeiro_complexity_2013}. However other studies have shown that for individuals with AD, those with a higher occupational attainment had a higher mortality rate than those with a lower occupation attainment \citep{Stern_Increased_1995,Stern_Rate_1999}. The caveat to previous studies assessing the effect of occupation on survival is that most consisted of populations with clinically diagnosed AD subjects, which can be unreliable \citep{beach_accuracy_2012}. Due to the inaccuracy of clinical diagnosis of AD, autopsy-confirmation is used for a definitive diagnosis \citep{grossman_mental_2016}. Without an accurate diagnosis of AD, any estimates of factors affecting survival are not reliable. In this paper, we aim to get improved estimates of the effect of occupation on survival from an autopsy-confirmed AD sample, adjusting for both truncation and dependence.

The remainder of this paper is organized as follows. In Section 2 we introduce the proposed EM method, including the estimation procedure and the large sample properties of the resulting estimators. In Section 3, we conduct a simulation study to assess the finite sample performance of the proposed estimators under dependent truncation. In Section 4, we apply the proposed method to estimate the effect of occupation on survival in individuals with autopsy-confirmed AD. Discussion and concluding remarks are given in Section 5. Proofs of the large sample results are outlined in the Appendix.

\section{Methods}  
We first introduce notation and assumptions. Let $T$ denote the survival time of interest (e.g. survival time from disease symptom onset), $L$ denote the left truncation time (e.g. time from disease symptom onset to entry into the study), $R$ denote the right truncation time (e.g. time from disease symptom onset to the end of study date), and $\bZ$ denote a $p\times 1$ vector of covariates. Let $N$ denote the size of the target population -- the population that would have been observed had there been no truncation present in the study. Due to double truncation, we only observe $(T_i,L_i,R_i,\bZ_i)$ for $i=1,...,n\leq N$ individuals who live long enough to enter the study (i.e. $T \geq L$) and do not live past the end of the study (i.e. $T\leq R$). Here we have denoted the population random variables from the target population without subscripts, and the sampling random variables from the observed sample with subscripts.

The proportional hazards model \citep{cox_regression_1972} is considered the standard regression model for analyzing traditional right-censored survival data. The model assumes that the covariate-specific hazard function is given by $\lambda_{\bZ}(t) = \lambda(t)\exp(\boldsymbol{\beta}'\boldsymbol{Z})$, where $\bbeta$ is a $p\times 1$ regression parameter vector, and $\lambda(t)$ is the baseline hazard function and is unspecified. When the survival data are subject to selection bias, Cox's partial likelihood approach \citep{cox_partial_1975} cannot be directly applied. This is because the observed data are not a representative sample of the target population, and therefore the observed, biased data do not follow the model that is assumed for the unbiased data from the target population. When the data is biased due to double truncation, the distribution of the observed survival time $T_i$ is given by:
\begin{align*}
P(T_i\leq t|\bZ_i) = P(T\leq t|\bZ_i,L\leq T\leq R) = \frac{P(T\leq t, L\leq T\leq R|\bZ_i)}{P(L\leq T\leq R|\bZ_i)} \neq P(T\leq t|\bZ_i),
\end{align*}
which differs from the distribution of the survival time $T$ from the target population. Therefore the resulting estimates of the regression coefficients based on data from the observed sample will be biased estimators of the regression coefficients from the target population. 

Under the assumption of independence between the survival and truncation times, \citep{rennert_cox_2017,shen_pseudo_2017,mandel_inverse_2017} adjust for double truncation by estimating the probability that a subject with survival time $T_i$ is observed, defined by $\widehat{\pi}_i = \widehat{P}(L\leq T\leq R|T=T_i)$, $i=1,...,n$. These probabilities are then used as weights or offsets in the Cox regression model. For example, under double truncation and independence between the survival times and truncation times, Rennert and Xie (2017) consistently estimate the true $p\times 1$ regression coefficient vector $\boldsymbol{\beta}_0$ by $\bbetaw$, the solution to 
\begin{align}\label{ScoreW}
\boldsymbol{U_w}(\boldsymbol{\beta},\widehat{\boldsymbol{\pi}}) &= \sum_{i=1}^n\int_0^{\tau}\widehat{\pi}_i^{-1}\bigg\{\boldsymbol{Z}_i(t) - \frac{\sum_{j=1}^n \widehat{\pi}_j^{-1}Y_j(t)\exp\{\boldsymbol{\beta}'\boldsymbol{Z}_j(t)\}\boldsymbol{Z}_j(t)}{\sum_{j=1}^n \widehat{\pi}_j^{-1}Y_j(t)\exp\{\boldsymbol{\beta}'\boldsymbol{Z}_j(t)\}}\bigg\}dN_i(t) = \boldsymbol{0},
\end{align}
where $\widehat{\boldsymbol{\pi}} = (\widehat{\pi}_1,...,\widehat{\pi}_n)$, $Y_i(t) = I(T_i \geq t)$, $N_i(t) = I(T_i \leq t)$, and $I$ is the indicator function. Here $\tau$ is the maximum of the observed event times. The standard Cox regression estimator \citep{cox_partial_1975} which ignores double truncation, $\bbetas$, is the solution to $\boldsymbol{U_w}(\boldsymbol{\beta},\boldsymbol{1}) = \boldsymbol{0}$, where $\boldsymbol{U_w}(\boldsymbol{\beta},\boldsymbol{1})$ is the score equation from the standard Cox model. 

The caveat of the approaches which adjust for double truncation is that they require estimating the distribution of the truncation times, which is needed to obtain the estimator of the selection probabilities $\widehat{\boldsymbol{\pi}}$. Existing methods to estimate the truncation distribution require independence between the survival and truncation times. When this independence assumption is violated, the estimator of the truncation distribution will be biased, and therefore the estimator of the selection probabilities $\widehat{\boldsymbol{\pi}}$ will be biased. Because the methods in \citep{rennert_cox_2017,shen_pseudo_2017,mandel_inverse_2017} depend on $\widehat{\boldsymbol{\pi}}$, the resulting regression coefficient estimators will also be biased. The severity of this bias is demonstrated by the simulation studies in the next section. 

When the survival times are conditionally independent of the truncation times given the covariate $\bZ$, the likelihood of the observed survival times, conditional on the truncation times and covariates, is given by
\begin{align*} 
L_{n}(\bbeta,\Lambda) = \prod_{i=1}^n\frac{\lambda(T_i)\exp(\boldsymbol{\beta}'\boldsymbol{Z}_i)\exp\{-\Lambda(T_i)\exp(\boldsymbol{\beta}'\boldsymbol{Z}_i)\}}{\alpha_i(\bbeta,\lambda)},
\end{align*}  
where $\alpha_i(\bbeta,\lambda) = \exp\{-\Lambda(L_i)\exp(\boldsymbol{\beta}'\boldsymbol{Z}_i)\} - \exp\{-\Lambda(R_i)\exp(\boldsymbol{\beta}'\boldsymbol{Z}_i)\}$ and $\Lambda(t) = \int_0^t\lambda(u)du$. That is, $\alpha_i(\bbeta,\lambda) = P(L\leq T\leq R|\bZ_i,L_i,R_i;\bbeta,\lambda)$ is the probability of observing a random subject from the target sample with covariate $\bZ_i$ and truncation times $L_i$ and $R_i$. Conditioning on the truncation times allows us to utilize the information in the covariates $\bZ$ to relax the assumption of independence to an assumption of conditional independence. Furthermore, this conditioning completely avoids the need to estimate the distribution of the truncation times.

The log-likelihood function, $\log L_n(\bbeta,\Lambda)$, can be expressed as
\begin{align} \label{eq:L}
l_n(\bbeta,\Lambda) = n^{-1}\sum_{i=1}^n\Big[\int_0^{\tau}\{\log\lambda(t)+\boldsymbol{\beta}'\boldsymbol{Z}_i-\Lambda(t)\exp(\boldsymbol{\beta}'\boldsymbol{Z}_i)\}dN_i(t)-\log\alpha_i(\bbeta,\lambda)\Big].
\end{align}  
Due to the difficulties of maximizing the log-likelihood (\ref{eq:L}) over all absolutely continuous cumulative hazard functions, we allow the estimator of $\lambda$ to be discrete. Because the maximum likelihood estimation (MLE) of $\boldsymbol{\beta}$ and $\lambda$ may be computationally intractable if directly solving the score equations for (\ref{eq:L}), we estimate $\bbeta_0$ and $\lambda$ using an EM algorithm. This has the advantage that its maximization step (M-step) only involves the complete-data likelihood. Based on the EM algorithm, we provide a convenient estimation approach to obtain estimators of the regression coefficients and baseline hazard function under left, right, or double truncation. This approach allows the survival and truncation times to be dependent through the covariate vector $\bZ$. Furthermore, it does not require the estimation of the truncation time distribution. The estimation approach given here can easily be implemented using standard software for the Cox regression model. 

\subsection{Proposed EM Algorithm}
Motivated by the approach in \citep{qin_maximum_2011}, who proposed EM algorithms for length-biased and right-censored data, Shen and Liu \citep{shen_pseudo_2017} proposed an EM algorithm to obtain pseudo MLEs of the regression coefficients from the Cox model under independent left and right truncation. They referred to their MLEs as pseudo because their proposed likelihood included the plug-in value of the estimator of the selection probabilities $\widehat{\boldsymbol{\pi}}$. However, as the authors point out, the estimated selection probabilities will be biased if the truncation times depend on the covariates $\textbf{Z}$. Hence, the resulting pseudo MLEs of the regression coefficients from the Cox model will also be biased.

We propose an EM algorithm for obtaining the MLE of $(\bbeta,\blambda)$ based on (\ref{eq:L}). This allows us to relax the assumption of independence required by the methods in \citep{rennert_cox_2017,shen_pseudo_2017,mandel_inverse_2017} to an assumption of conditional independence by avoiding the estimation of the truncation distribution (and corresponding selection probabilities). Similar to the approaches of \citep{shen_pseudo_2017} and \citep{qin_maximum_2011}, we let $t_1 <  ... < t_d$ denote the ordered, distinct failure times for $\{T_1,...,T_n\}$. We develop the EM algorithm based on the discrete version of $\Lambda$, which we redefine as a step function only taking jumps at $t_1,...,t_d$. Specifically, we set $\Lambda(t) = \sum_{t_j\leq t}\lambda_j$, were $\lambda_j$ is the positive jump at time $t_j$ for $j=1,...,n$. 

Our observed data consists of $\boldsymbol{O} = \{\bO_1,...,\bO_n\}$, where $\boldsymbol{O}_i \equiv (T_i,L_i,R_i,\boldsymbol{Z}_i)$ for $i=1,...,n$. Let $\boldsymbol{O}^* = \{T_{ir}^*;\text{ }i=1,...,n,r=1,...,m_i\}$ denote the truncated latent data, where $T_{ir}^*$ is the missing survival time for a subject with truncation times $(L_i,R_i)$ and covariate vector $\boldsymbol{Z}_i$ for $i=1,...,n$ and $r=1,...,m_i$. For notational convenience, we set $\boldsymbol{\theta} = (\boldsymbol{\beta},\boldsymbol{\lambda})$ and define the density of $T$ at time $t_j$, given $\boldsymbol{Z}_i$, as $f_i(t_j;\btheta) = \lambda_j\exp(\boldsymbol{\beta}'\boldsymbol{Z}_i)\exp\{-\sum_{s=1}^j\lambda_s\exp(\boldsymbol{\beta}'\boldsymbol{Z}_i)\}$, where $\blambda = (\lambda_{1},...,\lambda_{d})$. Assuming the latent survival times $T_{ir}^*$ take their values in $\{t_1,...,t_d\}$, the complete data log-likelihood is given by
\begin{align*}
l_{full}(\btheta;\bO,\bO^*) = \sum_{j=1}^d\sum_{i=1}^n\big[I(T_i=t_j) + \sum_{r=1}^{m_i}I(T_{ir}^* = t_j)\big]\log f_i(t_j;\boldsymbol{\beta},\boldsymbol{\lambda})
\end{align*}

To estimate the parameter $\btheta$, the EM algorithm begins by choosing an initial value for $\btheta$, say $\btheta^{(0)}$. In our setting, we can choose $\btheta^{(0)} = (\bbeta_s,\blambda_s)$, which are the estimates from the standard Cox model. For $k=0,1,2,...$, the expectation step (E-step) consists of calculating the expected value of the complete data log likelihood function $l_{full}(\btheta;\bO,\bO^*)$ with respect to the missing data $T_{ir}^*$, $i=1,...,n,r=1,...,m_i$, conditional on the observed data $(T_i,L_i,R_i,\bZ_i)$, $i=1,...,n$, under the current estimate $\btheta^{(k)}$. That is, we compute:
\begin{align*}
Q(\boldsymbol{\theta};\boldsymbol{\theta}^{(k)}) = E_{\boldsymbol{\theta}^{(k)}}\big[l_{full}(\btheta;\bO,\bO^*)\big|\bO\big]
\end{align*} 
In the maximization step (M-step), we choose $\btheta^{(k+1)}$ to maximize $Q(\boldsymbol{\theta};\boldsymbol{\theta}^{(k)})$. That is, we set
\begin{align*}
\btheta^{(k+1)} = \text{arg max}_{\btheta}Q(\boldsymbol{\theta};\boldsymbol{\theta}^{(k)})
\end{align*}
The E- and M-steps are carried out again, but this time with $\btheta^{(k)}$ replaced by $\btheta^{(k+1)}$. The E- and M-steps are then alternated repeatedly until $\|\btheta^{(k+1)}-\btheta^{(k)}\|<\epsilon$ for some prespecified error $\epsilon>0$.

The EM algorithm described here is under the double truncation setting. If only left truncation is present, the algorithm is easily adjusted by setting $R_i = \infty$ for $i=1,...,n$. When only right truncation is present, we set $L_i = -\infty$ for $i=1,...,n$. Note that when only left truncation is present, the standard Cox regression estimator can account for dependent left truncation by adjusting the risk set at a given time point to include all individuals who are alive and in the study at that time \citep{klein_survival_2003}. We denote this estimator by $\widehat{\bbeta}_{s,l}$. We show through simulations in the next section that when only left truncation is present, our proposed estimator and $\widehat{\bbeta}_{s,l}$ yield nearly identical results. 

\subsubsection{E-step}
At the $k^{th}$ iteration, define $\boldsymbol{\theta}^{(k)} =  (\boldsymbol{\beta}^{(k)},\boldsymbol{\lambda}^{(k)})$. Then,
\begin{align*}
Q(\boldsymbol{\theta};\boldsymbol{\theta}^{(k)}) & =  E_{\boldsymbol{\theta}^{(k)}}\Big[\sum_{j=1}^d\sum_{i=1}^n\big[I(T_i=t_j) + \sum_{r=1}^{m_i}I(T_{ir}^* = t_j)\big]\log f_i(t_j;\btheta)\big|\boldsymbol{O}\Big] \\
& = \sum_{j=1}^d\sum_{i=1}^n \Big\{I(T_i=t_j) + \sum_{j=1}^d\sum_{i=1}^nE_{\boldsymbol{\theta}^{(k)}}\Big[\sum_{r=1}^{m_i}I(T_{ir}^* = t_j)\Big|\boldsymbol{O}_i\Big]\Big\}\log f_i(t_j;\btheta) \\
& = \sum_{j=1}^d\sum_{i=1}^n \big\{I(T_i=t_j) + \sum_{j=1}^d\sum_{i=1}^nE_{m_i}\big[m_i\times E_{\bthetak}[I(T_{ir}^* = t_j)|\boldsymbol{O}_i]\big]\big\}\log f_i(t_j;\btheta),
\end{align*}
where
\begin{align*}
 E_{\bthetak}[I(T_{ir}^* = t_j)|\boldsymbol{O}_i] & =  P_{\bthetak}(T_{ir}^*=t_j|L_i,R_i,\bZ_i) = P_{\bthetak}(T=t_j|L_i,R_i,\bZ_i,\{T<L\}\cup\{T>R\}) \\
& = \frac{P_{\bthetak}(T=t_j,\{T<L\}\cup\{T>R\}|L_i,R_i,\bZ_i)}{P_{\bthetak}(\{T<L\}\cup\{T>R\}|L_i,R_i,\bZ_i)} = \frac{f_i(t_j;\bthetak)\times[I(t_j<L_i)+I(t_j>R_i)]}{1-\alpha_i(\bthetak)}.
\end{align*}
Since $m_i$ is the number of missing/truncated subjects with covariate values $\bZ_i$ and truncation times $L_i$ and $R_i$, $m_i$ follows a geometric distribution with success rate $\alpha_i(\btheta)$. Therefore when $\btheta = \bthetak$, $E[m_i] = \frac{1-\alpha_i(\bthetak)}{\alpha_i(\bthetak)}$. 

The complete data log likelihood is then given by 
\begin{align*}
Q(\btheta;\bthetak) = \sumjd\sumin\big\{I(T_i=t_j)+\frac{I(t_j<L_i)+I(t_j>R_i)}{\alpha_i(\bthetak)}f_i(t_j;\bthetak)\big\}\log f_i(t_j;\btheta)
\end{align*}
\subsubsection{M-step}
Let $\wijk=I(T_i=t_j)+\frac{I(t_j<L_i)+I(t_j>R_i)}{\alpha_i(\bthetak)}f_i(t_j;\bthetak)$. The complete data log likelihood can be written as 
\begin{align*}
Q(\btheta;\bthetak) = \sumjd\sumin\wijk\log f_i(t_j\btheta) = \sumjd\wplusjk\lambda_j + \sumin\wiplusk\bbeta'\bZ_i + \sumin\sumjd\sum_{s=1}^j\wijk\exp(\bbeta'\bZ_i)\lambda_s,
\end{align*}
where $\wplusjk=\sum_{i=1}^n\wijk$ and $\wiplusk=\sum_{j=1}^d\wijk$.

Treating $\wijk$ as constant, we set $\frac{\partial Q(\btheta;\bthetak)}{\partial\lambda_j} = 0$ to get a closed form solution to $\lambda_j$ as a function of $\bbeta$:
\begin{align}\label{eq:lambda0}
\lambda_j = \frac{\wplusjk}{\sum_{s=j}^d\sumin w_{is}^{(k)}\exp(\bbeta'\bZ_i)}, \quad j=1,...,d.
\end{align}
Differentiating $Q(\btheta;\bthetak)$ with respect to $\bbeta$ yields  
\begin{align*}
\frac{\partial Q(\btheta;\bthetak)}{\partial\bbeta} = \sumin\wiplusk\bZ_i + \sumin\sumjd\sum_{s=1}^j\wijk\bZ_i\exp(\bbeta'\bZ_i)\lambda_s.
\end{align*}
Setting the equation above equal to $0$ and inserting the equation for $\lambda_j$ yields
\begin{align}\label{eq:scoreEM}
\sumin\wiplusk\bZ_i - \sum_{s=1}^dw_{+s}^{(k)}\bigg\{\frac{\sumin\sum_{j=s}^d\wijk\bZ_i\exp(\bbeta'\bZ_i)}{\sumin\sum_{j=s}^d\wijk\exp(\bbeta'\bZ_i)}\bigg\} = \boldsymbol{0}.
\end{align}
The estimating equation (\ref{eq:scoreEM}) can be solved by specifying the ``weights'' option in the ``coxph'' function in R. First, a weight vector of length $nd$ must be created: $\boldsymbol{w}_{nd}^{(k)} = (w_{11}^{(k)},...,w_{1d}^{(k)},...,w_{n1}^{(k)},...,w_{nd}^{(k)})$. The corresponding failure time data and covariate vectors are also created with length $nd$ as follows: $\boldsymbol{T}_{nd} = (t_1,...,t_d,...,t_1,...,t_d)$ and $\bZ_{nd} = (\bZ_1,...,\bZ_1,...,\bZ_n,...,\bZ_n)$. Letting $\boldsymbol{\Delta}_{nd}$ be the identity vector of length $nd$, the solution to (\ref{eq:scoreEM}), which we denote by $\bbeta^{(k+1)}$, can be obtained with the following command: 
\begin{align*}
\text{coxph(Surv(}\boldsymbol{T}_{nd},\boldsymbol{\Delta}_{nd}\text{)} \sim \boldsymbol{Z}_{nd}\text{, weights = } \boldsymbol{w}_{nd}^{(k)}\text{, subset = which(} \boldsymbol{w}_{nd}^{(k)}\text{>0))}.
\end{align*}
Plugging $\bbeta^{(k+1)}$ into (\ref{eq:lambda0}) yields an updated estimator for $\blambda$, $\blambda^{(k+1)}$. We then set $\boldsymbol{\theta}^{(k+1)} =  (\boldsymbol{\beta}^{(k+1)},\boldsymbol{\lambda}^{(k+1)})$, and repeat the E- and M-steps. We continue to alternate between the E- and-M steps until $\|\boldsymbol{\theta}^{(k+1)}  - \boldsymbol{\theta}^{(k)}\| < \epsilon$, for some prespecified error $\epsilon > 0$. The MLE of the hazard ratio is then given by $\bbetaem = \boldsymbol{\beta}^{(k+1)}$. We denote the corresponding baseline hazard by $\blambdahat_{em} = \boldsymbol{\lambda}^{(k+1)}$, and the cumulative baseline hazard function by $\Lambdaemhat(t)=\sum_{t_j\leq t}\lambda_j^{(k+1)}$. 

The EM algorithm presented here falls into the general scheme of the ECM algorithm, and therefore its convergence to the local maximizer is guaranteed by the same conditions required for convergence of the ECM algorithm \citep{qin_maximum_2011}. The uniqueness of the resulting estimators are guaranteed by the regularity conditions in Appendix A.1.

\subsection{Asymptotic Properties}
In this section, we establish the strong consistency and asymptotic normality of the proposed EM estimators. Here we denote the proposed estimators by $\widehat{\boldsymbol{\theta}} = (\bbetaem,\Lambdaemhat)$, and denote the true regression coefficients and cumulative baseline hazard function $\btheta_0 = (\bbeta_0,\Lambda_0)$. The asymptotic properties of the proposed estimators refer to the situation when the total number of observed (non-truncated) subjects $n\rightarrow\infty$. The following theorems assume that the regularity assumptions in Appendix A.1 hold.

\textbf{Theorem 1:} Under the regularity assumptions given in the Appendix, $\bthetahat$ is consistent: As $n\rightarrow\infty$, $\bbetaem$ converges to $\bbeta_0$, and $\Lambdaemhat(t)$ converges to $\Lambda_0(t)$ almost surely and uniformly in $t$ for $t\in [0,\tau]$. 

The existence and uniqueness of the MLE can be proved based on the log-likelihood function 
\begin{align*} \label{eq:ln}
l_n(\bbeta,\blambda) = n^{-1}\sum_{i=1}^n\Big[&\int_0^{\tau}\bbeta'\bZ_idN_i(t) + \sum_{s=1}^d\log\lambda_s\times I(T_i=t_s) - \sum_{t_s\leq t_i}\lambda_s\exp(\bbeta'\bZ_i) \\
- & \log\Big\{\exp\Big(-\exp(\bbeta'\bZ_i)\sum_{t_s<L_i}\lambda_s\Big) - \exp\Big(-\exp(\bbeta'\bZ_i)\sum_{t_s\leq R_i}\lambda_s\Big)\Big\} \Big].
\end{align*}  
Theorem 1 can then be proved by applying the classical Kullback-Leibler information approach as in \citep{qin_maximum_2011}.

\textbf{Theorem 2:} Under the regularity assumptions given in the Appendix, $\sqrt{n}(\bthetahat-\btheta_0)$ converges weakly to a tight mean-zero Gaussian process.

Theorem 2 is proved using the Z-theorem for infinite dimensional equations \citep{vaart_weak_2000}. The proofs of Theorem 1 and Theorem 2 are outlined in Appendix A.2 and A.3, respectively. 

To obtain an estimate of the standard deviation of $\bbetaem$, we apply the simple bootstrap technique. In our setting, the bootstrap sample is obtained by drawing $n$ independent vectors $(T_j^b,L_j^b,R_j^b,\bZ_j^b)$, $j=1,...,n$, from the observed data vectors $(T_i,L_i,R_i,\bZ_i)$, $i=1,...,n$, with replacement. These data vectors are then used to obtain an estimate of regression coefficients, denoted by $\bbetaem^{(b)}$. This process is repeated $B$ times to obtain the $B$ estimators $\bbetaem^{(1)},....,\bbetaem^{(B)}$. The estimate of the standard deviation of $\bbetaem$ is computed by taking the standard deviation of the $\bbetaem^{(b)}$, $b=1,...,B$. We denote this estimate by $\widehat{\boldsymbol{\sigma}}_{\bbetaem}$. We show through simulation studies in the next section that the standard deviation of $\bbetaem$ is accurately estimated by $\widehat{\boldsymbol{\sigma}}_{\bbetaem}$.

\section{Simulations}
In this section we examine the performance of the proposed estimator under dependent truncation. We compare our proposed estimator to the weighted estimator which adjust for double truncation but assumes independence between the survival and truncation times. We also compare the proposed estimator to the estimator from the standard Cox regression model. In all simulations, the survival times were generated from a proportional hazards model with hazard function $\lambda(t)\exp(\beta_1Z_1+\beta_2Z_2)$, and follow a Weibull distribution with scale parameter $\nu = 0.001$ and shape parameter $\kappa = 5$. We set $\beta_1=\beta_2 = 1$, and generated the risk factors $Z_1$ and $Z_2$ from independent Unif[0,5] distributions. The truncation times were also simulated from Weibull distributions with scale parameter $\nu = 0.001$ and shape parameter $\kappa = 5$. The left truncation times were generated from a proportional hazards model with hazard function $\lambda_{L}(l)\exp(\beta_{L1}Z_1+\beta_{L2}X)$, and the right truncation times were simulated from a proportional hazards model with hazard function $\lambda_{R}(r)\exp(\beta_{R1}Z_1+\beta_{R2}Y)$. Here $X$ and $Y$ were generated from independent Unif[0,5] distributions, with $\beta_{L2} = \beta_{R2} = 1$. To adjust the proportion of missing data due to left and right truncation, the truncation times were multiplied by constants $c_l$ and $c_r$, respectively. A higher value of $c_l$ induced a higher proportion of missing data due to left truncation, while a lower value of $c_r$ induced a higher proportion of missing data due to right truncation. Because the survival, left, and right truncation times are all functions of $Z_1$ for $\beta_1\neq 0$, $\beta_{L1}\neq 0$, and $\beta_{R1}\neq 0$, they are dependent. However, the survival and truncation times are \textit{conditionally independent} given $Z_1$. To adjust the degree of dependence between $T$ and $L$, and $T$ and $R$, we varied the regression coefficients $\beta_{L1}$ and $\beta_{R1}$, respectively.  

We conducted 1000 simulation repetitions with sample sizes of $n$ = 100 and 250. To obtain $n$ observations after truncation, we simulated $N=\frac{n}{1-q}$ observations, where $q$ is the proportion of truncated data. For each simulation, we estimated $\bbeta=(\beta_1,\beta_2)$, using the proposed EM estimator $\bbetaem=(\betaemone,\betaemtwo)$, the weighted Cox regression estimator $\bbetaw=(\betawone,\betawtwo)$, and the standard Cox regression estimator $\bbetas=(\betasone,\betastwo)$. Of the estimators which adjust for double truncation under the independence assumption, we only focus on $\bbetaw$ from \citep{rennert_cox_2017}, as previous simulations (not shown here) have concluded that this estimator and that in \citep{mandel_inverse_2017} are nearly identical, and both outperform the estimators in \citep{shen_pseudo_2017}. For each estimator, we calculated the estimated bias, observed sample standard deviations (SD), estimated standard errors ($\widehat{\textrm{SE}}$), and the average empirical coverage probability of the 95\% confidence intervals (Cov). To compare the efficiency of the estimators which adjust for double truncation to the efficiency of the standard estimator, we calculated the relative mean-squared error (MSE) of $\widehat{\beta}_j$ to $\betasj$, $j=1,2$. That is, we computed rMSE$(\widehat{\beta}_j) = \frac{\text{MSE}(\widehat{\beta}_j)}{\text{MSE}(\betasj)}$ for $\widehat{\beta}_j = \betaemj$ and $\widehat{\beta}_j = \betawj$. We used 200 bootstrap resamples to estimate the standard error of $\betaemj$ and $\betawj$, $j=1,2$.

Table 1 shows the results of the simulations described above. In the first model, we set $\beta_{L1} = -1$ to induce a negative dependence between the survival times and left truncation times, which resulted in a correlation of -0.35. In the second model, we set $\beta_{L1} = 0$ to induce independence between the survival and left truncation times. In the third model, we set $\beta_{L1} = 1$ to induce a positive dependence between the survival times and left truncation times, which resulted in a correlation of 0.35. Here $c_l$ and $c_r$ were chosen such that 25\% of the survival times were left truncated and 25\% of the survival times were right truncated, which resulted in $q\approx$ 0.50. The parameter $\beta_{R1}$ was set to $1$ in all models, which resulted in a correlation of 0.35 between the survival times and right truncation times.    

In all models, the proposed EM estimators $\betaemone$ and $\betaemtwo$ had little bias, while the standard estimators $\betasone$ and $\betastwo$ were biased. The weighted estimator $\betawone$ was heavily biased in all models, while $\betawtwo$ was biased in the first set of models ($\rho_{LT} = -0.35$). The observed sample standard deviations of the proposed estimators were accurately estimated by the bootstrap technique, and the coverage probabilities of the proposed estimators were all close to the nominal level of 0.95. The coverage probabilities of $\betasone$ and $\betastwo$ were well below the nominal level, as were the coverage probabilities for $\betawone$. Furthermore, the mean-squared errors of the proposed estimators were lower than those of the weighted and standard estimators in almost all settings, indicating that the proposed EM method is more efficient.

We further explored the bias and MSE of these estimators as a function of left and right truncation proportion (Figure 2). We set $\beta_{L1} = \beta_{R1} = 1$ and $n=250$, which corresponded to the setting of the last model in Table 1, inducing a positive dependency between the survival times and both left and right truncation times. The proposed estimators had little bias, regardless of truncation proportion. Even under mild truncation, the weighted estimator $\betawone$ of the regression coefficient corresponding to $Z_1$, which is correlated with the truncation times, was biased. This bias increased drastically as the proportion of right truncation increased. The bias was relatively small for both the proposed and weighted estimator of the regression coefficient corresponding to $Z_2$, which is uncorrelated with the truncation times. Both standard estimators $\betasone$ and $\betastwo$ were heavily biased in this setting. The MSE of $\betaemj$ was significantly lower than the MSE of $\betawj$ for $j=1,2$. Furthermore, in most cases the MSE of $\betaemj$ was lower than the MSE of the standard estimator $\betasj$, i.e. rMSE($\betaemj$) $< 1$ for $j=1,2$. 

In Figure 3, we compared the bias and MSE of these estimators under varying truncation proportions, when the assumption of independence holds (i.e. $\beta_{L1} = \beta_{R1} = 0$). The proposed EM estimators $\betaemj$ and weighted estimators $\betawj$ had little bias, while the standard estimators $\betasj$ were biased for $j=1,2$. We also compared the rMSE of $\betaemj$ and $\betawj$ to $\betasj$, $j=1,2$. As indicated by the bottom row of Figure 3, the proposed EM estimators had similar efficiency to the weighted estimators when the independence assumption holds. When the proportion of missing data due to left and right truncation were approximately equal, the standard estimator was more efficient than the proposed EM estimator and the weighted estimator. This is because the bias due to left truncation canceled out with the bias due to right truncation when these proportions were equal, which yielded a lower MSE. 

The standard Cox regression model can accommodate left truncation when the left truncation time is conditionally independent of the survival times given the observed risk factors. We compare the estimator from this model to our proposed estimator under dependent left truncation only in Appendix A.4. As shown in Figure 4, the results are nearly identical, and both estimators outperform the weighted estimators in this setting.

\section{Application to Alzheimer's Disease}
We illustrate our method by considering an autopsy-confirmed AD study conducted by the Center for Neurodegenerative Disease Research at the University of Pennsylvania. The target population for the research purposes of this study consists of all subjects with AD symptom onset before 2012 that met the study criteria and therefore would have been eligible to enter the center. Our observed sample contains all subjects who entered the center between 1995 and 2012, and had an autopsy performed before July 1, 2012. Thus one criterion for a subject to be included in our sample is that they did not succumb to AD before they entered the study, yielding left truncated data. In addition, our sample only contains subjects who had an autopsy-confirmed diagnosis of AD, and therefore we have no knowledge of subjects who live past the end of the study. Thus our data is also right truncated. Our data consists of n=91 subjects, all of whom have event times. The event time of interest is the survival time ($T$) from AD symptom onset. The left truncation time ($L$) is the time between the onset of AD symptoms and entry into the study (i.e. initial clinic visit). The right truncation time ($R$) is the time between the onset of AD symptoms and the end of the study, which is taken to be July 1, 2012. Due to double truncation, we only observe subjects with $L\leq T\leq R$.

We are interested in assessing the effect of occupation on survival in AD. Occupation is often used as a proxy for cognitive reserve (CR), which hypothesizes that individuals develop cognitive strategies and neural connections throughout their life times through experience such as occupation, education, and other forms of mental engagement \citep{Valenzuela_Assessment_2007}. A common hypothesis in the literature is that CR has protective role in the brain and modulates the effects of AD because of compensatory strategies obtained from a higher level of professional performance and therefore lengthens survival during the course of the disease \citep{Sanchez_Study_2011,Ientile_Survival_2013}.  

However some studies have shown a higher mortality rate in AD individuals with higher occupational attainment \citep{Stern_Increased_1995,Stern_Rate_1999}. This supports an alternative theory of CR; individuals with higher CR tolerate more pathology which delays the onset of the disease. Because higher age of AD symptom onset is associated with an increased risk of mortality, this would support the hypothesis that those with higher CR would have an increased risk of mortality. There are two caveats to the studies described above. The first is that these studies consisted of populations with clinically diagnosed AD subjects, which can be unreliable. The second caveat is that the statistical analyses were subject to confounding, since age of AD symptom onset was not recorded nor adjusted for. 

Here we are interested in obtaining improved estimates of the effect of occupation on survival from an autopsy-confirmed cohort of individuals with AD who have a known age of disease symptom onset. We use the highest occupational attainment for a given subject as a proxy for their CR. Primary occupation was classified and ranked based on the US census categories. In the following analyses, subjects who were classified as manager, business/government, and professional/technical workers were labeled as having \textit{high occupational attainment} in our study. Subjects classified as unskilled/semiskilled, skilled trade or craft, and clerical/office workers were classified as having \textit{low occupational attainment}. This classification is consistent with previous studies \citep{Massimo_Occupational_2015,Stern_Increased_1995,massimo_occupational_2018}. Age at AD symptom onset was estimated based on a family report at first contact with the individual. 

We first check the assumption of independence between the observed survival and truncation times using the conditional Kendall's tau proposed by Martin and Betensky \citep{martin_testing_2005}. The resulting p-value is 0.038, and therefore we reject this independence assumption. The corresponding Kendall's tau statistic is $\tau_K = (0.20,0.16)$, indicating positive dependence between the survival times and truncation times. The positive dependence between the left truncation times and survival times is clinically plausible because doctors often attribution the symptoms of early onset AD (onset of AD before 65 years of age) to other causes such as depression and stress, hence delaying the study entry time. Since younger age at onset is also associated with higher survival, this induces a positive dependence between the left truncation times and survival times.

Due to the dependence between the survival and truncation times, we apply the proposed method to estimate the effect of occupation on survival, adjusting for age at AD symptom onset and sex. Table 2 displays the results from the Cox regression model using the proposed EM estimators, weighted estimators, and the standard estimators. Using the proposed method, the estimated log hazard ratio for age at AD symptom onset is 0.029 (p-value = 0.016), indicating that AD individuals who have symptom onset one year later are roughly 3\% more likely to die than subjects who have symptom onset a year earlier ($e^{0.029}=1.03$). The estimated effect of female is -0.636 (p-value = 0.023), indicating that males are almost twice as likely to die than females ($e^{0.636}=1.89$). These effects are nearly doubled using the weighted method which assumes independence, however the effects are not statistically significant (p-values = 0.117 and 0.088, respectively). 

High occupational attainment is associated with increased survival in all models. Under the proposed method, the effect of high occupational attainment on survival is -0.673 (p-value = 0.009), indicating that those with a low occupational attainment are approximately twice as likely to die than those with a high occupational attainment ($e^{0.673}=1.96$). This effect is attenuated under the weighted and standard methods, and neither method yielded statistically significant estimates (p-values are 0.186 and 0.158, respectively).  

\section{Discussion}
We proposed a novel method which relaxes the independence assumption between the observed survival and truncation times in the Cox model under left, right, or double truncation to an assumption of conditional independence between the observed survival and truncation times. We obtained consistent and asymptotically normal estimators of the regression coefficients and baseline hazard function by maximizing the conditional likelihood of the observed survival times using an EM algorithm. The simulation studies confirmed that the proposed estimators had little bias in small samples, while the na\"{\i}ve estimators from the Cox models which ignore truncation or assume independence were biased. The existing methods which adjust for truncation but assume independence resulted in heavily biased estimators of the regression coefficients for risk factors of survival that were also correlated with the truncation times. Furthermore, the proposed estimators were more efficient than the na\"{\i}ve estimators in most of the simulation settings. 

We applied our proposed method to an autopsy-confirmed sample of individuals with Alzheimer's disease (AD). AD is a major neurodegenerative disease which currently affects 5.3 million people in the United States according to the Alzheimer's Association. In 2017 alone, AD and other dementias will have cost the nation an estimated \$259 billion. Autopsy-confirmation is needed for a definitive diagnosis of AD, and a definitive diagnosis is necessary to accurately estimate the effect of potential risk factors associated with a given neurodegenerative disease. However, autopsy-confirmed samples of neurodegenerative diseases are subject to an inherent selection bias due to double truncation. Existing methods which adjust the Cox model in the presence of double truncation assume that the observed survival and truncation times are independent. This assumption may not be reasonable for studies of neurodegenerative diseases. In our data example, this independence assumption was rejected. Therefore, previous methods are not appropriate for our setting.

Given the severity of Alzheimer's disease on patients, their caregivers, and society, accurate estimation of the effects of risk factors on survival is crucial. One such factor, cognitive reserve (CR), is hypothesized to lengthen survival during the course of the disease. Using occupation as a proxy for CR, we estimated the effect of CR on survival in an autopsy-confirmed AD sample. Using our proposed method to adjust for both left and right truncation and dependence between the survival and truncation times, we found that a low occupational attainment was associated with shortened survival. Compared to existing methods, the estimated hazard ratios for occupation on survival were larger under our proposed method. This is consistent with many studies concluding that an individual's occupation may provide a protective effect and lengthen survival in AD. These findings suggest the importance of incorporating occupation in treatment trials and prognostic considerations in individuals with AD. 

A limitation of our proposed method is that it in its current form, it cannot properly handle time-varying covariates measured after study entry, such as cognitive test scores. This is a consequence of the estimation procedure, which uses an expectation-maximization algorithm to estimate the latent survival times conditional on the observed truncation times and risk factors. This leads to predicting survival times based on risk factors measured after death for those missing subjects whose survival time is less than their left truncation time, which may yield biased regression coefficient estimators. 

The proposed method has useful implications for observational studies. Double truncation has been shown to be present in a variety of studies, such as studies of clinically diagnosed Parkinson's disease \citep{mandel_inverse_2017}, childhood cancer \citep{moreira_semiparametric_2010}, astronomy data \cite{efron_nonparametric_1999}, and studies based on registry data \citep{shen_pseudo_2017,bilker_semiparametric_1996}. In fact, any data pulled from a disease registry will be subject to inherent right truncation, since data is only recorded for subjects who have the disease and are entered in the registry by the time the data is extracted \citep{bilker_semiparametric_1996}. In certain cases, the data will also be subject to left truncation \citep{shen_pseudo_2017,bilker_semiparametric_1996}. In a similar fashion, studies which only include data from individuals whose event times fall within the time course of the study are subject to double truncation \citep{moreira_semiparametric_2010}. Therefore careful consideration of the study design must be taken into account when fitting the Cox proportional hazards model. Furthermore, the assumption of independence should always be tested, given the high sensitivity of existing methods to this assumption. For example, a quick application of a Kendall's conditional Tau test \citep{martin_testing_2005} revealed this independence assumption is violated in the AIDS data used in \cite{shen_pseudo_2017}. We therefore recommend using the proposed estimators in most practical settings, since they have little bias, and in most situations, have a lower mean-squared error compared to existing estimators under left, right, or double truncation, under a wide range of dependence structures.

\section*{Appendix}
We first define some notation. Let $\bbetahat_n=\bbetaem$ and $\widehat{\Lambda}_n = \widehat{\Lambda}_{em}(\cdot)$, and set $\bthetahatn=(\bbetahatn,\Lambdahatn)$. The true parameter is denoted by $\btheta_0 = (\bbeta_0,\Lambda_0)$.

A.1 Regularity Assumptions
\begin{enumerate}
\item The true hazard function $\lambda_0(\cdot)$ is continuously differentiable, $\Lambda_0(0) = 0$, and $\Lambda_0(\tau) < \infty$.
\item The true parameter vector $\bbeta_0$ lies in a compact set $\mathbb{B}$. The set $\mathbb{A}$ contains all nondecreasing functions $\Lambda$ satisfying regularity assumption 1.
\item $E\|Z\|$ and $E\|\exp(\bbeta'\bZ)\|$ are bounded, where $\|\bz\|\equiv \sqrt{z_1^2+...+z_p^2}$.
\item The information matrix $-\partial^2E[l_n(\bbeta,\widehat{\blambda}(\bbeta))]/\partial\bbeta^2|_{\bbeta=\bbeta_0}$ is positive definite. Here $\widehat{\blambda}(\bbeta) = \blambda_{em}$ is used to emphasize the dependence on $\bbeta$.
\item If $P(\boldsymbol{b}'\bZ = c) = 1$ for some constant $c$, then $\boldsymbol{b}=0$. 
\end{enumerate}
Assumptions 1 and 2 are required for stochastic approximation. Assumptions 3 and 4 are needed to establish the asymptotic properties of the regression parameter estimates from the Cox model \citep{andersen_statistical_1997}. Assumption 5 implies no covariate colinearity and thus ensures that the model is identifiable.

A.2 Consistency: Proof of Theorem 1

Since each function of $\blambda$ in $l_n(\bbeta,\blambda)$ is concave or strictly concave, and the summation of concave functions is concave, the log-likelihood function $l_n(\bbeta,\blambda)$ is strictly concave in $\blambda$. Therefore we can find a unique maximizer $\widehat{\blambda}(\cdot,\bbeta)$ of $l_n(\bbeta,\blambda)$ for each $\bbeta$ in a compact set $\mathbb{B}$. The existence of the NPMLE for $(\bbeta,\blambda)$ follows by compactness of $\mathbb{B}$ for the likelihood $l_n(\bbeta,\widehat{\blambda}(\cdot,\bbeta))$, which is continues in $\bbeta$. Uniqueness is guaranteed by Assumption for in A.1 for large samples.  

Here we show that if $\bthetahatn$ converges, it must converge to $\btheta_0$. As $\bthetahat_n$ maximizes the log-likelihood given in (\ref{eq:L}), $\lntheta$, the empirical Kullback-Leibler distance $\lnthetahat-\lnthetazero$ must be nonnegative. Suppose $\bthetahatn$ converges to some $\btheta^* = (\bbeta^*,\Lambda^*)$. Then by the strong law of large numbers (SLLN), $\lnthetahat-\lnthetazero$ must converge to the negative Kullback-Leibler distance between $P_{\btheta^*}$ and $P_{\btheta_0}$. Here $P_{\btheta}$ is the probability measure under the parameter $\btheta$. Since the Kullback-Leibler distance and $\lnthetahat-\lnthetazero$ are nonnegative, the Kullback-Leibler distance between $P_{\btheta^*}$ and $P_{\btheta_0}$ must be zero. Therefore $P_{\btheta^*}=P_{\btheta_0}$ almost surely, and it then follows from model identifiability that $\btheta^* = \btheta_0$. Therefore if $\bthetahatn$ converges, it must converge to $\btheta_0$.

The technical details to show that $\bthetahatn$ indeed converges are similar to those in \citep{Murphy_consistency_1994}. The idea is to find a further convergent subsequence for any subsequence of $\bthetahatn$, and then apply Helly's selection theorem. Here we provide only a sketch of the proof. The first step is to show that $\bthetahatn$ stays bounded. By regularity assumption 3, $\bbetahatn$ is in a compact set and is therefore bounded. To show $\Lambdahatn$ is bounded, we make use of the fact that the empirical Kullback-Leibler distance $l_n(\bthetahatn) - l_n(\bar{\btheta})$ is always non-negative for each $\bar{\btheta}$ in the parameter set. Using the approach of Murphy (1994), it can be shown that if $\Lambdahatn$ does indeed diverge to $\pm\infty$, then it is possible to construct some sequence $\bar{\btheta}_n$ such that $l_n(\bthetahatn) - l_n(\bar{\btheta})$ eventually becomes negative infinity, which contradicts the nonnegativity of the empirical Kullback-Leibler distance. 

Since $\bthetahatn$ stays bounded, we can apply Helly's selection principal to find a further convergent subsequence $\bthetahat_{n_k} = (\bbetahat_{n_k},\widehat{\Lambda}_{n_k})$ for any subsequence of $\bthetahatn$ indexed by $\{1,...,n\}$. By the classical Kullback-Leibler information approach, and the SLLN, $\bthetahat_{n_k}$ must converge to $\btheta_0$. It then follows from Helly's selection theorem that the entire sequence $(\bbetahatn,\Lambdahatn(t))$ must converge to $(\bbeta_0,\Lambda_0(t))$ for every $t\in[0,\tau]$, where $\tau=t_d$ is the maximum of the observed event times. Since $\Lambda_0(\cdot)$ is assumed to be monotone and continuous, the convergence of $\Lambdahatn(t)$ is uniformly in $t\in[0,\tau]$. Because the proof is carried out for a fixed $\omega$ in the underlying probability space $\Omega$, where the SLLN is applied countably many times, the convergence here is also almost surely a true convergence.

A.3 Asymptotic Normality: Proof of Theorem 2

Here we outline the proof for the weak convergence of $\bthetahat_n$, which follows the proof for weak convergence in \citep{qin_maximum_2011}. The proof consists of the application of empirical process theory and the Z-theorem for infinite dimensional estimating equations \citep{vaart_weak_2000}. 

Denote the score equation for $\bbeta$ by $\boldsymbol{U}_{1n}(\btheta) = \partial\lntheta/\partial\bbeta$. To obtain the score equation $\Lambda(\cdot)$, we define the submodel $d\Lambda_{\epsilon} = (1+\epsilon h)d\Lambda$, where $h$ is a bounded and integrable function. Setting $h(\cdot)=I(\cdot\leq t)$, the score equation for $\Lambda$ is given by
$\boldsymbol{U}_{2n}(t,\btheta) = \frac{\partial l_n(\bbeta,\Lambda_{\epsilon})}{\partial\epsilon}|_{\epsilon=0}$. 

We denote the vector of the score functions by $\boldsymbol{U}_n(\cdot,\btheta) = [\boldsymbol{U}_{1n}(\btheta) , \boldsymbol{U}_{2n}(t,\btheta)]$. The expectation $E_0$ under the true value $\btheta_0$ is given by $
\boldsymbol{U}_0(\cdot,\btheta) = [\boldsymbol{U}_{10}(\btheta) , \boldsymbol{U}_{20}(t,\btheta)]$, where $\boldsymbol{U}_{10}(\btheta) = E_0[\boldsymbol{U}_{1n}(\btheta)]$ and $\boldsymbol{U}_{20}(t,\btheta) = E_0[\boldsymbol{U}_{2n}(\cdot,\btheta)]$.

By the definition of the MLE, $\boldsymbol{U}_n(\cdot,\bthetahat_n)=0$. Since $\boldsymbol{U}_0(\cdot,\btheta_0)=0$, we can show that $|\sqrt{n}\{\boldsymbol{U}_0(\cdot,\bthetahat_n)-\boldsymbol{U}_n(\cdot,\bthetahat_n)\} -\sqrt{n}\{\boldsymbol{U}_n(\cdot,\btheta_0)-\boldsymbol{U}_0(\cdot,\btheta_0)\}|=o_p(1)$. The estimating equation evaluated at $\btheta_0$, $\sqrt{n}\boldsymbol{U}_n(\cdot,\btheta_0) = \sqrt{n}\{\boldsymbol{U}_n(\cdot,\btheta_0)-\boldsymbol{U}_0(\cdot,\btheta_0)\}$, is a sum of iid terms. We can therefore use empirical process theory to show that $\sqrt{n}\boldsymbol{U}_n(\cdot,\btheta_0)$ converges weakly to $\mathbb{W}=(\mathbb{W_1},\mathbb{W}_2)$, where $\mathbb{W}_1$ is a Gaussian random vector and $\mathbb{W}_2$ is a tight Gaussian process. The covariance matrix for $\mathbb{W}_1$ is given by $\boldsymbol{\Sigma}_{11} = E_0\{\boldsymbol{U}_{1n}(\btheta_0)^{\otimes 2}\}$, and the covariance between $\mathbb{W}_2(s)$ and $\mathbb{W}_2(t)$ is given $\Sigma_{22}(s,t) = E_0\{\boldsymbol{U}_{2n}(s,\btheta_0)\boldsymbol{U}_{2n}(t,\btheta_0)\}$.   

Applying the Z-theorem for the infinite dimensional estimating equations (theorem 3.3.1 in \citep{vaart_weak_2000}), we have that under the regularity conditions in A.1, $\sqrt{n}(\bthetahat_n-\btheta_0)$ converges weakly to a tight mean-zero Gaussian process $-\dot{U}_{\btheta_0}^{-1}(\mathbb{W})$. 

Here $\dot{U}_{\btheta_0}$ is the Fr\'echet derivative of the map   $\boldsymbol{U}_0(\cdot,\btheta)$ evaluated at $\btheta_0$. Using arguments similar to Appendix A.5 in \citep{qin_maximum_2011}, we can show $\boldsymbol{U}_0(\cdot,\btheta)$ is Fr\'echet differentiable and the Fr\'echet derivative, $\dot{U}_{\btheta_0}$, is continuously invertible. By definition of the Fr\'echet derivative, we have that $\dot{U}_{\btheta_0}\{\sqrt{n}(\bthetahat_n-\btheta_0)\} = -\sqrt{n}\{\boldsymbol{U}_n(\cdot,\btheta_0)-\boldsymbol{U}_0(\cdot,\btheta_0)\} + o_p(1)$. This completes the proof.

A.4 Simulations under dependent left truncation

Under dependent left truncation only, we compared our proposed method to the weighted method and the standard method which accounts for dependent left truncation. To adjust the correlation between the left truncation times and survival times, we varied the parameter $\beta_{L1}$ between $-1$ and $1$. In this setting, a value of $\beta_{L1}=0$, which yields a correlation of $0$, indicates independence between the left truncation times and survival times. We denote the standard regression coefficient estimator which adjusts for dependent left truncation as $\bbetahat_{s,l} = (\widehat{\beta}_{s,l,1},\widehat{\beta}_{s,l,2})$. As shown in Figure 4, $\betaemj$ and $\widehat{\beta}_{s,l,j}$ had little bias, while the weighted estimators $\betawj$ were biased for $j=1,2$. As indicated by the bottom row of Figure 4, the proposed EM estimators had similar efficiency to the standard estimators which accounted for dependent left truncation, and both estimators were more efficient than the weighted estimators. 

\section*{Acknowledgements}
We would like to thank Dr. Murray Grossman for his contribution to the clinical aspect
of this paper. Mr. Rennert received support from NIH National Institute of Mental Health
grant T32MH065218 and Dr. Xie from NIH grant R01-NS102324, AG10124, AG17586, and
NS053488.

\bibliography{ALLreferences}

\newpage
\begin{figure}[h!]
	\begin{center}
		\includegraphics[scale=0.775,clip,trim= 1.1cm 1.6cm 0 0cm]{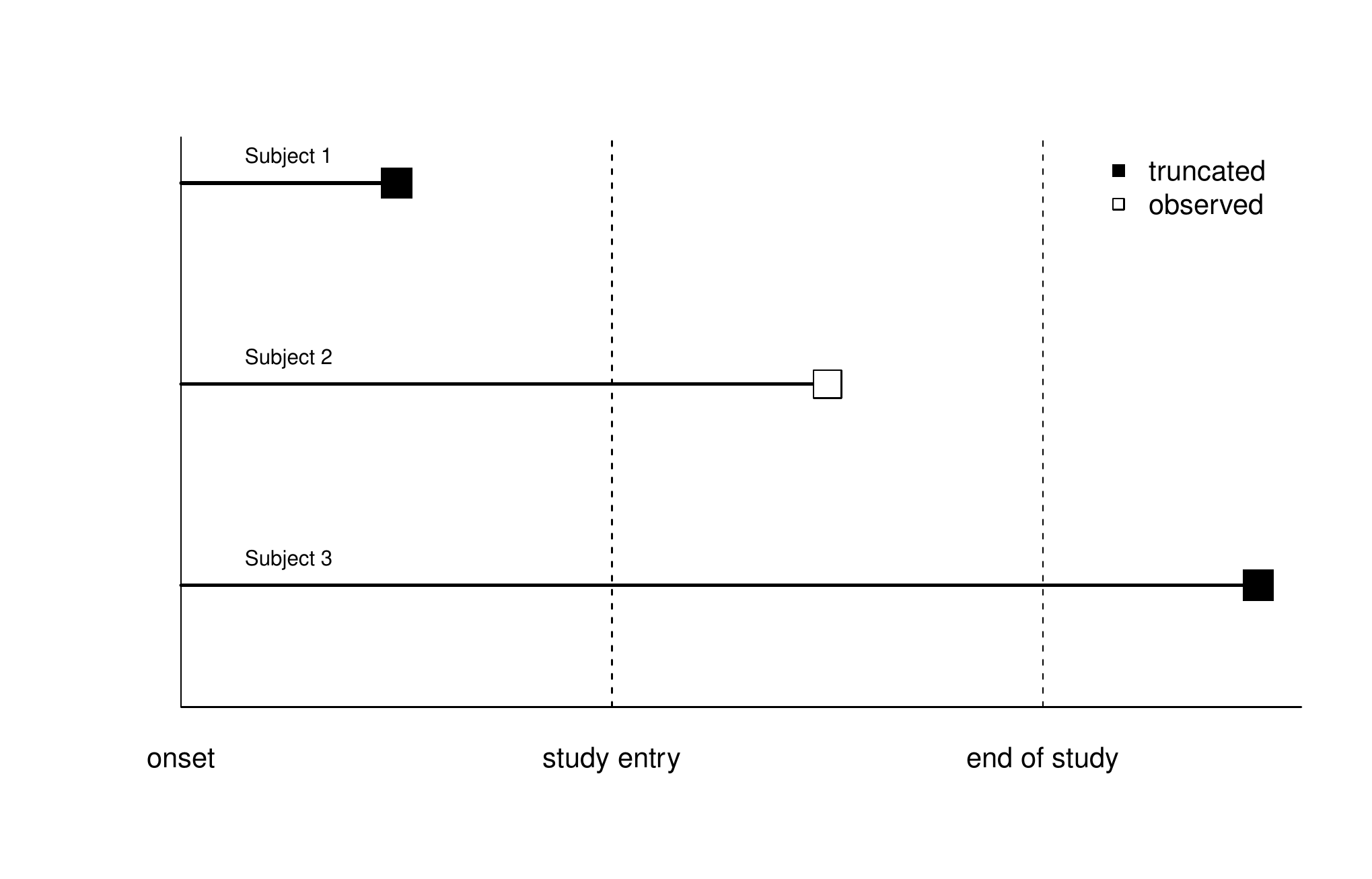}
	\end{center}
	\caption{\footnotesize In this hypothetical example, we assume subjects 1, 2, and 3 all have similar times of disease symptom onset. For illustrative purposes, we also assume that subjects 1, 2, and 3 have the same study entry time, however this need not be the case. Here the x-axis represents time, and the squares represent the terminating events. Subject 1 is left truncated because they die before they enter the study. Subject 2 enters the study and dies before the end of the study, and is therefore observed. Subject 3 is right truncated because they live past the end of the study, and therefore do not have an autopsy performed.}
	\label{DoubleTruncationPlot}
\end{figure}

\newpage
\begin{figure}[h!]
	\begin{center}
		\includegraphics[scale=0.775,clip,trim= 1cm 2.6cm 0 2.7cm]{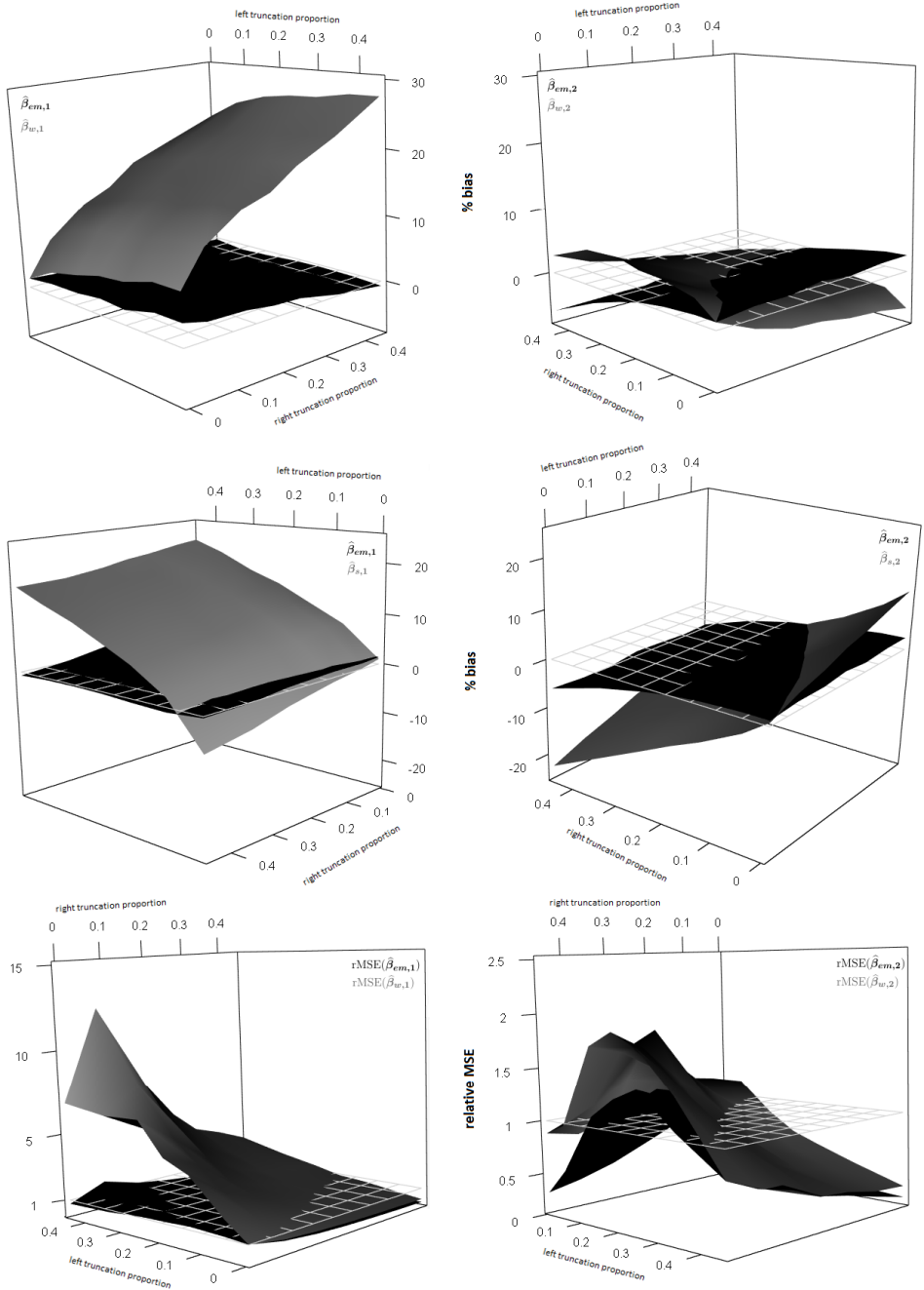}
	\end{center}
	\caption{\footnotesize Comparing bias and MSE (mean-squared error) of estimators across different left and right truncation proportions, under \textit{dependent} survival and truncation times. Survival times generated from proportional hazards model with hazard function $\lambda(t)\exp(\beta_1Z_1 + \beta_2Z_2)$, with $\beta_1=\beta_2=1$. Survival times conditionally independent of left and right truncation times given $Z_1$. For $j=1$ (left column) and $j=2$ (right column): Top row compares bias of proposed EM estimator $\widehat{\beta}_{em,j}$ (\textbf{black}) to weighted estimator $\widehat{\beta}_{w,j}$ (\textcolor{dark-gray}{\textbf{gray}}), which does not account for dependent truncation. Middle row compares bias of $\widehat{\beta}_{em,j}$ (\textbf{black}) to the standard estimator $\widehat{\beta}_{s,j}$ (\textcolor{dark-gray}{\textbf{gray}}), which ignores truncation completely. Bottom row compares rMSE($\widehat{\beta}_{em,j})$ (\textbf{black}) to rMSE($\widehat{\beta}_{w,j})$ (\textcolor{dark-gray}{\textbf{gray}}). Here rMSE($\widehat{\beta}$) =  $\frac{\text{MSE}(\widehat{\beta}_j)}{\text{MSE}(\widehat{\beta}_{s,j})}$ is the relative MSE of the estimator $\widehat{\beta}_j$ to the standard estimator $\widehat{\beta}_{s,j}$.}
	\label{plot11}
\end{figure} 

\newpage
\begin{figure}[h!]
	\begin{center}
		\includegraphics[scale=0.775,clip,trim= 1cm 2.6cm 0 2.7cm]{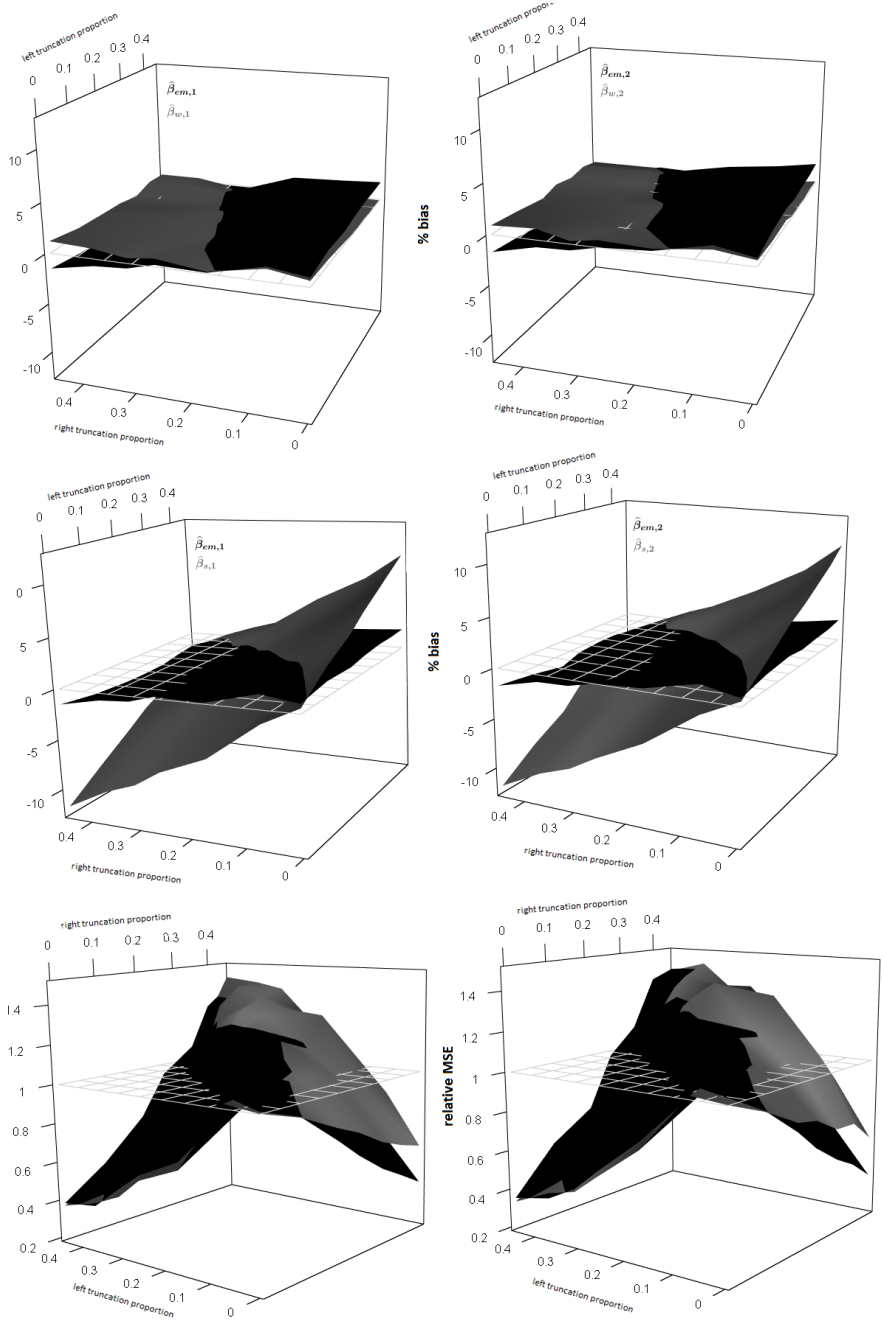}
	\end{center}
	\caption{\footnotesize Comparing bias and MSE (mean-squared error) of estimators across different left and right truncation proportions, under \textit{independent} survival and truncation times. Survival times generated from proportional hazards model with hazard function $\lambda(t)\exp(\beta_1Z_1 + \beta_2Z_2)$, with $\beta_1=\beta_2=1$. For $j=1$ (left column) and $j=2$ (right column): Top row compares bias of proposed EM estimator $\widehat{\beta}_{em,j}$ (\textbf{black}) to weighted estimator $\widehat{\beta}_{w,j}$ (\textcolor{dark-gray}{\textbf{gray}}), which does not account for dependent truncation. Middle row compares bias of $\widehat{\beta}_{em,j}$ (\textbf{black}) to the standard estimator $\widehat{\beta}_{s,j}$ (\textcolor{dark-gray}{\textbf{gray}}), which ignores truncation completely. Bottom row compares rMSE($\widehat{\beta}_{em,j})$ (\textbf{black}) to rMSE($\widehat{\beta}_{w,j})$ (\textcolor{dark-gray}{\textbf{gray}}). Here rMSE($\widehat{\beta}$) =  $\frac{\text{MSE}(\widehat{\beta}_j)}{\text{MSE}(\widehat{\beta}_{s,j})}$ is the relative MSE of the estimator $\widehat{\beta}_j$ to the standard estimator $\widehat{\beta}_{s,j}$.}
	\label{plot00}
\end{figure} 
\newpage

\begin{figure}[h!]
	\begin{center}
		\includegraphics[scale=0.775,clip,trim= 1cm 2.6cm 0 2.7cm]{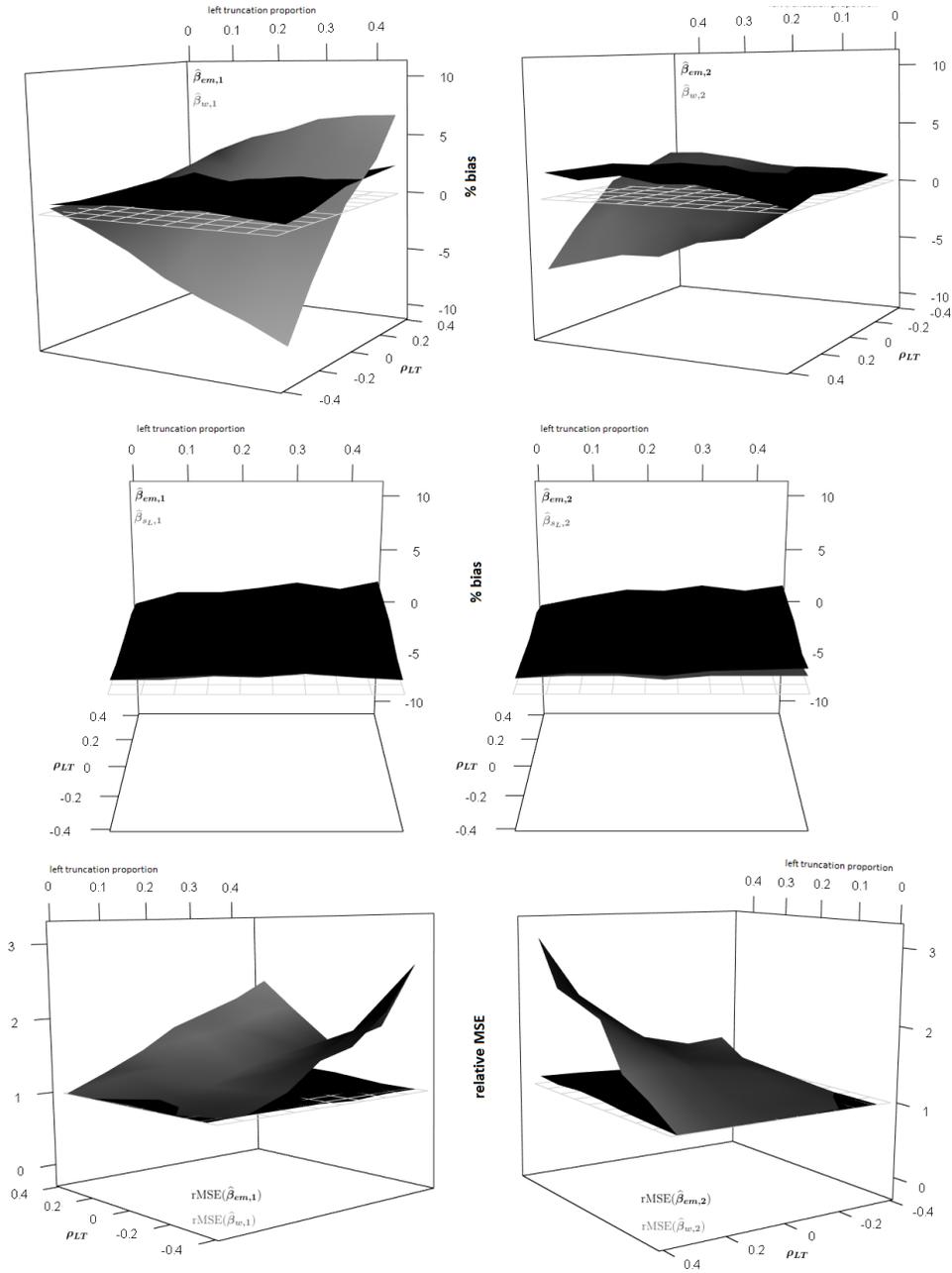}
	\end{center}
	\caption{\footnotesize Comparing bias and MSE (mean-squared error) of estimators under \textit{dependent left truncation}. Here $\rho_{LT}$ is the correlation between the left truncation and survival time.  Survival times generated from proportional hazards model with hazard function $\lambda(t)\exp(\beta_1Z_1 + \beta_2Z_2)$, with $\beta_1=\beta_2=1$. For $j=1$ (left column) and $j=2$ (right column): Top row compares bias of proposed EM estimator $\widehat{\beta}_{em,j}$ (\textbf{black}) to weighted estimator $\widehat{\beta}_{w,j}$ (\textcolor{dark-gray}{\textbf{gray}}), which does not account for dependent truncation. Middle row compares bias of $\widehat{\beta}_{em,j}$ (\textbf{black}) to the standard estimator under left truncation $\widehat{\beta}_{s_L,j}$ (\textcolor{dark-gray}{\textbf{gray}}), which accounts for dependent left truncation. Bottom row compares rMSE($\widehat{\beta}_{em,j})$ (\textbf{black}) to rMSE($\widehat{\beta}_{w,j})$ (\textcolor{dark-gray}{\textbf{gray}}). Here rMSE($\widehat{\beta}$) =  $\frac{\text{MSE}(\widehat{\beta}_j)}{\text{MSE}(\widehat{\beta}_{s_L,j})}$ is the relative MSE of the estimator $\widehat{\beta}_j$ to the standard estimator under left truncation $\widehat{\beta}_{s_L,j}$.}
	\label{plot00}
\end{figure} 
\newpage

\begin{table}[h!]
	\centering
	\footnotesize
	\caption{Simulation results} 
	{\em Here $\rho_{LT}$ is the correlation between the left truncation and survival time. The correlation between the right truncation and survival time is fixed at 0.35. The EM method produces the proposed estimator $\betaem$, which adjusts for double truncation and dependence. The weighted method produces the estimator $\widehat{\beta}_{w}$, which adjusts for double truncation, but assumes independence between the survival and truncation times. The standard method assumes no truncation and produces the estimator $\widehat{\beta}_s$, the solution to the standard Cox score equation. Here SD is the empirical standard deviation of estimates across simulations, $\widehat{\textrm{SE}}$ is the average of the estimated standard errors. For an estimator $\widehat{\beta}$, rMSE = $\frac{\text{MSE}(\widehat{\beta})}{\text{MSE}(\widehat{\beta}_s)}$, where MSE is the mean-squared error. Cov is the coverage of 95\% confidence intervals. Survival times generated from hazard function $\lambda(t)\exp(\beta_1Z_1 + \beta_2Z_2)$, with $\beta_1=\beta_2=1$. Survival times conditionally independent of left and right truncation times given $Z_1$.}\\
	\begin{tabular}{llr|ccccc|ccccc}
		\hline \hline
		\multicolumn{1}{c}{$\rho_{LT}$} & 
		Method & 
		\multicolumn{1}{c}{$n$} & 
		\multicolumn{1}{c}{Bias($\widehat{\beta}_1$)} & 
		\multicolumn{1}{c}{SD($\widehat{\beta}_1$)} & 
		\multicolumn{1}{c}{$\widehat{\textrm{SE}}$($\widehat{\beta}_1$)} &
		\multicolumn{1}{c}{rMSE($\widehat{\beta}_1$)} & 
		\multicolumn{1}{c}{Cov($\widehat{\beta}_1$)} &
		\multicolumn{1}{c}{Bias($\widehat{\beta}_2$)} & 
		\multicolumn{1}{c}{SD($\widehat{\beta}_2$)} & 
		\multicolumn{1}{c}{$\widehat{\textrm{SE}}$($\widehat{\beta}_2$)} &
		\multicolumn{1}{c}{rMSE($\widehat{\beta}_2$)} & 
		\multicolumn{1}{c}{Cov($\widehat{\beta}_2$)} \\ 
		\hline
		\multirow{6}{*}{-0.35}
		& EM & 100 & 0.01 & 0.13 & 0.14 & 1.05 & 0.96 & -0.00 & 0.14 & 0.13 & 0.82 & 0.94 \\ 
		& weighted & 100 & 0.10 & 0.15 & 0.16 & 1.94 & 0.95 & 0.06 & 0.15 & 0.15 & 1.11 & 0.95 \\ 
		& standard & 100 & -0.05 & 0.12 & 0.12 & 1.00 & 0.91 & -0.10 & 0.12 & 0.11 & 1.00 & 0.82 \\ \rule{0pt}{5ex}
		& EM & 250 & -0.01 & 0.08 & 0.08 & 0.64 & 0.95 & -0.02 & 0.08 & 0.08 & 0.38 & 0.94 \\ 
		& weighted & 250 & 0.08 & 0.09 & 0.09 & 1.54 & 0.89 & 0.04 & 0.09 & 0.09 & 0.55 & 0.92 \\ 
		& standard & 250 & -0.07 & 0.07 & 0.07 & 1.00 & 0.83 & -0.11 & 0.07 & 0.07 & 1.00 & 0.60 \\
		\hline
		\multirow{6}{*}{0.00}
		& EM & 100 & -0.01 & 0.12 & 0.13 & 0.67 & 0.96 & 0.00 & 0.13 & 0.13 & 1.07 & 0.96 \\ 
		& weighted & 100 & -0.05 & 0.12 & 0.13 & 0.75 & 0.94 & 0.02 & 0.12 & 0.13 & 0.99 & 0.96 \\ 
		& standard & 100 & -0.10 & 0.11 & 0.12 & 1.00 & 0.84 & -0.04 & 0.11 & 0.11 & 1.00 & 0.92 \\  \rule{0pt}{5ex}
		& EM & 250 & -0.01 & 0.08 & 0.08 & 0.38 & 0.95 & -0.01 & 0.08 & 0.08 & 0.80 & 0.95 \\ 
		& weighted & 250 & -0.05 & 0.08 & 0.08 & 0.56 & 0.89 & 0.01 & 0.07 & 0.08 & 0.76 & 0.95 \\ 
		& standard & 250 & -0.10 & 0.07 & 0.07 & 1.00 & 0.68 & -0.05 & 0.07 & 0.07 & 1.00 & 0.88 \\ 
		\hline 
		\multirow{6}{*}{0.35}
		& EM & 100 & 0.03 & 0.16 & 0.16 & 0.72 & 0.95 & 0.01 & 0.14 & 0.14 & 1.15 & 0.96 \\ 
		& weighted & 100 & 0.20 & 0.15 & 0.15 & 1.74 & 0.78 & 0.00 & 0.14 & 0.15 & 1.23 & 0.96 \\ 
		& standard & 100 & 0.13 & 0.13 & 0.12 & 1.00 & 0.82 & -0.05 & 0.12 & 0.12 & 1.00 & 0.92 \\  \rule{0pt}{5ex}
		& EM & 250 & 0.00 & 0.09 & 0.09 & 0.45 & 0.94 & -0.01 & 0.08 & 0.08 & 0.73 & 0.94 \\ 
		& weighted & 250 & 0.18 & 0.09 & 0.09 & 2.22 & 0.45 & -0.01 & 0.09 & 0.09 & 0.87 & 0.95 \\ 
		& standard & 250 & 0.11 & 0.08 & 0.07 & 1.00 & 0.67 & -0.06 & 0.07 & 0.07 & 1.00 & 0.84 \\ 
		\hline
	\end{tabular}
	\label{tab:Results}
\end{table}

\newpage

\begin{table}[h!]
	\normalsize
	\caption{Application results of Cox model. Event time is years from AD symptom onset to death.}
	\begin{threeparttable}
		\setlength{\tabcolsep}{6pt}
		\begin{tabular}{lcccccc}
			\hline
			& EM && Weighted && Unweighted &\\
			Predictor & $\betaem (\widehat{\textrm{SE}})$ & p-value &  $\betaw (\widehat{\textrm{SE}}$) & p-value &  $\betas (\widehat{\textrm{SE}})$ & p-value \\
			\hline
			Age onset &  0.029 (0.012) & 0.016  & 0.047 (0.030) & 0.117 & 0.035 (0.013) & 0.013\\
			Female & -0.636 (0.280) & 0.023  & -1.026 (0.602) & 0.088 & -0.532 (0.223) & 0.017\\
			High occupation &  -0.673 (0.257) & 0.009  & -0.464 (0.351) & 0.186 & -0.487 (0.345) & 0.158\\
			\hline
		\end{tabular}
	\end{threeparttable}
\end{table}

\end{document}